\documentstyle[amssymb,12pt,thmsa]{article}

\input{tcilatex}
\begin{document}

\begin{center}
\strut {\bf On Some General Physical Aspects of the Real Physical Space and
Time}

{\sl A.M. Gevorkian , R. A. Gevorkian* }

{\sl Institute of Physical Research, 378410, Ashtarak-2, Republic of Armenia}

{\sl *Institute of Radiophisics and Electronics, 378410, Ashtarak-2,
Republic of Armenia}

\strut

\smallskip {\bf Abstract}
\end{center}

{\it The paper discusses some aspects of real physical space and time:
structure of electrodynamics; structure of gravitational-inert theory in the
real world; formation of the real matter; Mach's principle and the
radiational cosmic mass; spread and transmission of information in the space
of the real world; fundamental and main physical constants, protostar\ or
3+1-dimensional matter; and some philosophical and general physical
approaches to solutions of physical problems.}

\smallskip

1. In the real 4-dimensional space the 3-dimensional and 1-dimensional
subspaces are distinguished with the coefficient of imaginary relativity $ci$%
\ between their dimensions. That is why it is thought of as 3+1-dimensional
and relativistic space.

2. Electrodynamics - To the 3-dimensional subspace, determined by
3-dimensional polar vector field (electrical), axial (magnetic) vector is
added in the form of a relativistic addition. The union of these vector
fields creates the polar-axial (electromagnetic) anti-symmetric tensor of
the fourth order. Lagrange's vector equation, which is obtained on the basis
of Lagrangian, becomes Maxwell's equations of electrodynamics.

For such a space, as a consequence, the communicational wave equation is
obtained which is determined by\ d'Alembert's operator

\begin{center}
\begin{equation}
\square A=\Delta A-\frac 1{c^2}\frac \partial {\partial t^2}A=0
\label{eq2.1}
\end{equation}
\end{center}

3. Earlier we showed that an equation is added to Einstein's equation
determined by the counter-projection of tensor [1]. Also an expression was
received which presented the geometric interpretation of the active
principle of equivalence:

\begin{equation}
\left( -\right) P_{ik}=P_{ik}  \label{eq2.2}
\end{equation}

Just like the magnetic field is added to the electric field in
electrodynamics, here the inert field is added to the gravitational field in
the form of equation $G_{ik}=2P_{ik}$\ . These fields are associated with
each other through the principle of active equivalence [1].

4. On the basis of the principle of active equivalence in [1] we received an
algebraically transcendental equation, the core of which is illustrated in
fig 1.

Fig. 1a presents the equilibrial main state of a virtual particle with real
mass equal to zero. On fig. 1b, after deformation of 3+1-dimensional space,
another equilibrial state is created, which determines the real, stable
matter with a finite mass. The surface of integration is determined from
point 1, in the direction of increase of $x$, as well as in the direction of
decrease of $1/x$. The mass obtained this way is the source of gravitational
and inert fields, and the electric charge is a necessary attribute during
formation of the usual matter [1]. However, it must be noted that the
divergence from the real value is approximately 10\%, which is due to the
spin of the object.

5. In [1] we received and formulated Mach's principle. $M_i=m_i+\overline{m}%
_i$\ is the effective mass effecting the test body. Relative to the test
body, on the leverage $i$\ on both sides, there will always be compensating
each other masses $m_i$\ and $\overline{m}_i$\ . Then the effective impact
on that leverage is equal to zero, and therefore $M_i=m_i+\overline{m}_i=0$.
Naturally, the summation by $i$\ is also equal to zero.

\begin{equation}
\sum M_i=0  \label{eq5.1}
\end{equation}

The average value of mass (3) is equal to zero, and the shift from the
average value is different from zero. Then $M=\left( \sum \overline{\Delta }%
m_i^2\right) ^{1/2}$represents the radiational cosmic mass. It creates
oscillations in the Metagalactic environment (vacuum oscillations). We think
that the real radiation (three-degree equilibrial radiation) is responsible
for this phenomenon.

6. The second law of thermodynamics is a typical device for converting
information into physical entropy. Information can be viewed as fuel for
perpetuum motion\ of the second type. Virtual particles, which are members
of non-excited environment, are in one main state, and the union of these
particles creates macroscopically non-excited vacuum. That is why entropy,
which is determined as the logarithm of statistical weight of macroscopic
states of subsystems, is equal to zero, $S=0$.

On the basis of\ Nernst's\ theorem, we conclude that the non-excited vacuum
can be viewed as a thermodynamic system at the absolute zero. For
information transmission it is necessary to excite the environment (vacuum),
by filling it with excited particles (create fluctuations).

7. The fundamental constants are $G$, $c$, $h$\ . Their combinations can
create $hc$, $c^4/G$, the fundamental charge and force respectively, formed
from the combination of two constants. In [1] the non-measure\ constant $%
\alpha ^2=\frac{hc}{e^2}$\ (with 10\% accuracy) is received, which
associates the real matter with the fundamental constant of space
(environment).

Let us consider energy $e^2/r$\ along the entire accessible length (from $r$%
\ to $R$). It is assumed that the integral represents the fundamental
charge, i.e.

\strut 
\[
hc=\int_{r_0}^R\frac{e^2}rdr 
\]

or

\begin{equation}
R=r_0\exp \frac{hc}{e^2}  \label{eq7.1}
\end{equation}

where $r_0$\ is the fundamental length, , and $r_0=\left( \frac{Gh}{c^3}%
\right) ^{1/2}$, $\ \exp \frac{hc}{e^2}\approx 10^{59\div 60}$\ is the
cosmological factor. Then the numerical value of $R$\ is estimated as $R\sim
10^{27}$sm.

The measures of the Metagalaxy are estimated at the same value through
astrophysical observations. If the fundamental length in the formula is
substituted by the fundamental value of mass, then we receive

\begin{equation}
M=m_0\exp \frac{hc}{e^2},\qquad \qquad M\sim 10^{55}gr
\end{equation}

The value of the mass of the Metagalaxy is also in accordance with
astrophysical estimations.

Inside the atom the integral of Coulomb\ potential is equal to $\left(
hc\right) ^{1/2}$, and is equal to zero outside of the atom.

\[
\left( hc\right) ^{1/2}=\int_{r_{kl}}^R\frac erdr\qquad or\qquad
R=r_{kl}\exp \frac{\left( hc\right) ^{1/2}}e,\qquad R\sim 10^{-10}sm 
\]

The hydrogen atom size is received when $r_{kl}=\frac{e^2}{mc^2}$\
(classical radius of the proton).

Another fundamental constant $c^4/G$\ represents Coulomb's, as well as
Newton's forces for fundamental physical values. It is assumed that after
formation of the Real World, Newton's force between the Metagalaxy and the
hydrogen atom, does not change on the distance equal to the classical radius
of the hydrogen atom.

Let us introduce $R=\frac{e^2}{\left( m_pm_e\right) ^{1/2}c^2}$\ - the value
of the classical hydrogen atom radius (analogous to that of charged
particles). For Newton's force between the Metagalaxy and the hydrogen atom
we receive:

\[
F=\frac{GM\left( m_pm_e\right) ^{3/2}c^4}{e^4}\qquad or\qquad \frac{c^4}G=%
\frac{GM\left( m_pm_e\right) ^{3/2}c^4}{e^4} 
\]

After substituting M with its value received from (5), we obtain:

\[
\frac{e^4}{G^2\left( m_pm_e\right) ^{3/2}}=\left( \frac{hc}G\right)
^{1/2}\exp \alpha ^2 
\]

where the dimendsionless value:

\begin{equation}
\frac{e^2}{Gm_pm_e}=\alpha ^{2/3}\exp 2/3\alpha ^2\sim 10^{-40\div 41}
\end{equation}

represents the ratio of Coulomb force to Newton's force in the hydrogen atom.

The numerical results received here surprisingly accurately coincide with
experimental data. Thus, main dimendsionless physical constants like $\alpha
^2=\frac{e^2}{hc},$\ $\frac{e^2}{Gm_pm_e}=\beta ^2=\alpha ^{2/3}\exp
2/3\alpha ^2$\ and cosmological constants like $R=r_0\exp \alpha ^2,$\ $%
M=m_0\exp \alpha ^2$\ - the size and mass of Metagalaxy, and Hubble's
constant

\begin{equation}
H=\frac cR=\frac c{r_0\exp \alpha ^2}
\end{equation}

as well as the radiational width

\begin{equation}
{\normalsize \Delta \nu =}\frac 43\frac \sigma {hc}{\normalsize T}^4%
{\normalsize R}
\end{equation}

are added to fundamental physical constants $h,$\ $c,$\ $G$, which describe
the physical environment. They are also added to such fundamental physical
values as

\[
r_0=\left( \frac{Gh}c\right) ^{1/2},\qquad t_0=\frac{Gh}{c^5},\qquad
l_0=hc,\qquad F_0=\frac{c^4}{G^5} 
\]

which are formed by the fundamental physical constants.

According to Stefan-Boltzmann's law, the energy density in each point of the
real world is determined by a three-degree equilibrial radiation (relic
radiation):

\begin{equation}
E=\frac{4\sigma }cT^4
\end{equation}

where $\sigma $\ is Stefan-Boltzmann's constant $\sigma =\frac{\pi ^2k^2}{%
60\eta ^3c^2}$.

The energy along the length of the atom (e.g. hydrogen atom) is equal to:

\[
E=\frac 43\frac \sigma cT^4R 
\]

where $R$\ is thedimendsionless radius (length) of the atom.

The radiational width of spectral lines in atoms is determined by the
constant radiation of the environment

\[
E=h\Delta \nu =\frac 43\frac \sigma cT^4R 
\]

and for the radiational width in specters we have

\begin{equation}
\Delta \nu =\frac 43\frac \sigma {hc}T^4R\sim 10^8
\end{equation}

We also have the cosmologial factor

\begin{equation}
A=\exp \alpha ^2\sim 10^{60\div 61}
\end{equation}

and the atomic factor

\begin{equation}
a=\exp \alpha =10^5
\end{equation}

8. On the basis of generalization of the theory of non-stationar starts, and
the analysis of activity of the nuclei of galaxies, and galaxies with blue
excess, allowed V. Ambartsumian to conclude that the process of star
formation and non-stationar state of some formations are connected with
protostar substance [2], which degrades and becomes usual matter. This
process is accompanied by release of enormous amount of energy. There have
not been any direct observations of objects with protostar substance.

We think that such protostar substance may be the $3+1$-dimensional
substance, which decomposes into smaller unstable fragments until reaching
the state of stable particles (electron, proton, photon), releasing energy.
Then these particles form the usual 3-dimensional substance, the most basic
of which is the hydrogen atom. Stable configurations, such as white dwarfs,
neutron stars (Oppenheimer, Volkov), hyperonic stars (Ambartsumian, Saakian,
Vartanian) etc, can not be protostar objects. However, there is one
interesting work by Muradian [3,4], which describes the morphology of the
macroscopic hadron (two-dimensional heavy hadron), which, due to being
unusual (application of nuclear physics laws in astrophysics), may possibly
be the protostar substance.

If the proton and the electron are the stable ''relic'' substance ($3+1$%
-dimensional), then in order to understand the protostar macroscopic
substance, it is necessary to build the internal electrodynamics of the
charged particle. The physical presentation of their microscopic forms
(electron and proton), may serve as protostar macroscopic substance (one
possibility is the ball lightning). Therefore it is necessary to build the
electrodynamics of charged particles, or receive Maxwell's equations
(non-traditional), which describe the internal structure of electromagnetic
field.

9. a) Let us consider equality

\begin{equation}
P+\overline{P}=1
\end{equation}

The sum of probabilities of a phenomenon and its counter-phenomenon is equal
to one.

Example: n number of shooters have a hypothetical probability of to hit the
target $p_i$. It is necessary to find the probability of hitting the target
in the case of simultaneous shooting. Due to factorization of the
counter-probability

\begin{equation}
\overline{P}=\prod_i^n\left( 1-p_i\right) ,we\text{ }receive\quad
P=1-\prod_i^n\left( 1-p_i\right) \qquad
\end{equation}

Problem is solved.

This method allows to solve if not all, then the majority of problems in the
theory of probabilities and statistics (for big n).

b) Definition

\[
f\left( \overline{x}\right) =\overline{f}\left( x\right) 
\]

If a function with a counter-argument becomes its own counter-function, such
a function is called a counter-odd function.

\[
f\left( \overline{x}\right) +\overline{f}\left( x\right) =1 
\]

One of the particular solutions of the functional equation is

\[
f\left( x\right) =\sin ^2x,\qquad x+\bar{x}=\frac \pi 2,\qquad x+\bar{x}=1 
\]

The simultaneous happening of the phenomenon and the counter-phenomenon
takes place in anti-phase. If we imagine the electric and magnetic fields to
be the phenomenon and the counter-phenomenon respectively, then Maxwell's
electrodynamics can be presented through the model of the example.

c) In [1] we received

\[
G_{ik}+\overline{G}_{ik}=2P_{ik} 
\]

where $P_{ik}$\ is the tensor of counter-projection, and for scalars the
tensor equation becomes equation (14). \smallskip 
\[
G+\overline{G}=1 
\]

We received that the sum of scalar curvature $G$\ and curvature $\overline{G}
$\ is equal to one. In the framework of this model we received the equation
of the inert field.

Cases a) and c) can be brought to case b) if we assume that the object and
the counter-object change and depend on the parameter of state. More
examples can be brought which are described by this model. Based on these
examples, the general law can be formulated.

The core of the law of dialectics on ''struggle and unity of opposites''
lies not so much in the ''struggle'', rather it lies in their unity. This
means that in the process of the ''struggle'', one opposition becomes
enriched at the account of the other one. Thus, oppositions struggle, and in
this process they turn into each other either partially or completely. These
changes happen in the anti-phase. The process of opposing may be stationar,
like in examples a) and c), or non-stationar, as in b). If we assume that
objects described in examples a) and c) are non-stationar, then they also
can be brought to example b). Electrodynamics is described by the model of
example b), where the object is the electric filed, and the counter-object
is the magnetic field. Maxwell's equations form a unified system where the
electric and the magnetic fields oppose to each other and change in the
anti-phase. As a consequence of the unified system the wave equation is
received.

\begin{equation}
\square A=0
\end{equation}

If we assume that in examples a) and c) the objects are non-stationar, then
for the first example the wave equation of amplitudes of probabilities is
expected, while for the third example the wave equation of the symmetric
tensor $g_{ik}$\ is expected.

\begin{equation}
\square g_{ik}=0
\end{equation}

Similarly to equations for 4-dimensional vector potential, the
electromagnetic field can also be generalized into a 10-component tensor
equation like (16). Rules of d'Alambertian effect on tensor are not yet
completely clear. It is necessary that construction $\square $\ be some kind
of general differential operator of the second order, which contains a
member like

\[
\frac{\partial ^2g_{ik}}{\partial x^\mu \partial x^\upsilon }=0 
\]

This operator requires further mathematical work. In [1] we used extreme
values of the object and the counter-object and received two pairs of
equivalent equations:

\[
G_{ik}+2\frac{\left( g_{ik}-\delta _{ik}\right) }{\xi _i\xi ^i}=0,\quad 
\overline{G}_{ik}=0,\quad First\text{ }equivalent\text{ }pair 
\]

\smallskip 
\[
\overline{G}_{ik}-2\frac{\left( g_{ik}-\delta _{ik}\right) }{\xi _i\xi ^i}%
=0,\quad G_{ik}=0,\quad Second\text{ }equivalent\text{ }pair 
\]

which describes gravitational and inert fields.

\smallskip Examples used in point 9 demonstrate that the object and the
counter-object coexist and form a unified system. The real physical space
and time are described by such general physical aspects.

\begin{center}
\smallskip \quad

\smallskip

\smallskip

\smallskip

\smallskip

\smallskip

Literature
\end{center}

1. A. M. Gevorkian, R.A. Gevorkian. ''Inertia in the Structure of
Four-dimensional Space''. Automated e-print archives at arXiv.org: Physics,
General Relativity and Quantum Cosmology, gr-qc/0002046, 14.2.2000
xxx.lanl.gov.

2. V. A. Ambarcumyan ''About Protostars'' Reports of Academy of Science
(DAN), 16, 97, 1953

3. R. M. Muradian ''Astrophysics'' 11, 237, 1975

4. R. M. Muradian ''Astrophysics'' 13, 63 1975

\end{document}